\documentclass[a4paper,twocolumn,showpacs]{revtex4-1}
\usepackage{graphicx}
\usepackage{epstopdf}
\usepackage{dcolumn}
\usepackage{bm}
\usepackage{float}
\usepackage[                   
            pdfstartview=FitH,
            colorlinks, 
            pdfborder=001,   
            linkcolor=blue,
            anchorcolor=blue,
            citecolor=blue,
            urlcolor=blue
            ]{hyperref}
\usepackage[utf8]{inputenc}
\pdfoutput=1
\begin{document}
\title{Andreev reflection adjusting in the multi-terminal device with the kink states}
\author{Lin Zhang$^{1,3}$, Chao Wang$^{*1,2}$, Peipei Zhang$^{1}$, and Yu-Xian Li$^{*1}$}

\affiliation{$^{1}$College of Physics and Hebei Advanced Thin Film Laboratory,
Hebei Normal University, Shijiazhuang 050024, People's Republic of China\\
$^{2}$College of Physics, Shijiazhuang University, Shijiazhuang 050035, People's Republic of China\\
$^{3}$Shool of Mathematics and Science, Hebei GEO University, Shijiazhuang 050031, People's Republic of China}

\begin{abstract}
At the domain wall between two regions with the opposite Chern number, there should be the one-dimensional chiral states, which are called as the kink states. The kink states are robust for the lattice deformations.  We design a multi-terminal device with the kink states to study  the local Andreev reflection and the crossed Andreev reflection.  In the three-terminal device the local Andreev reflection can be suppressed completely  for either $\varepsilon_{0}=0.5\Delta_{s}$ and $V_{0}=0.1t$ or $\varepsilon_{0}=-0.5\Delta_{s}$ and $V_{0}=-0.1t$, where $\varepsilon_{0}$ is the on-site energy of the graphene terminals and $V_{0}$ is the stagger energy of the center region.  The coefficient of the crossed Andreev reflection can reach $1$ in the four-terminal device. Besides adjusting the phase difference between superconductors,  the local Andreev reflection and the crossed Andreev reflection can be controlled by  changing the on-site energy and the stagger energy in the four-terminal device. Our results give some new ideas to design the quantum device in the future.

\end{abstract}


\date{\today}

\maketitle

\section{Introduction}
\begin{figure}[t]
  \centering
  \includegraphics[width=1.0\columnwidth]{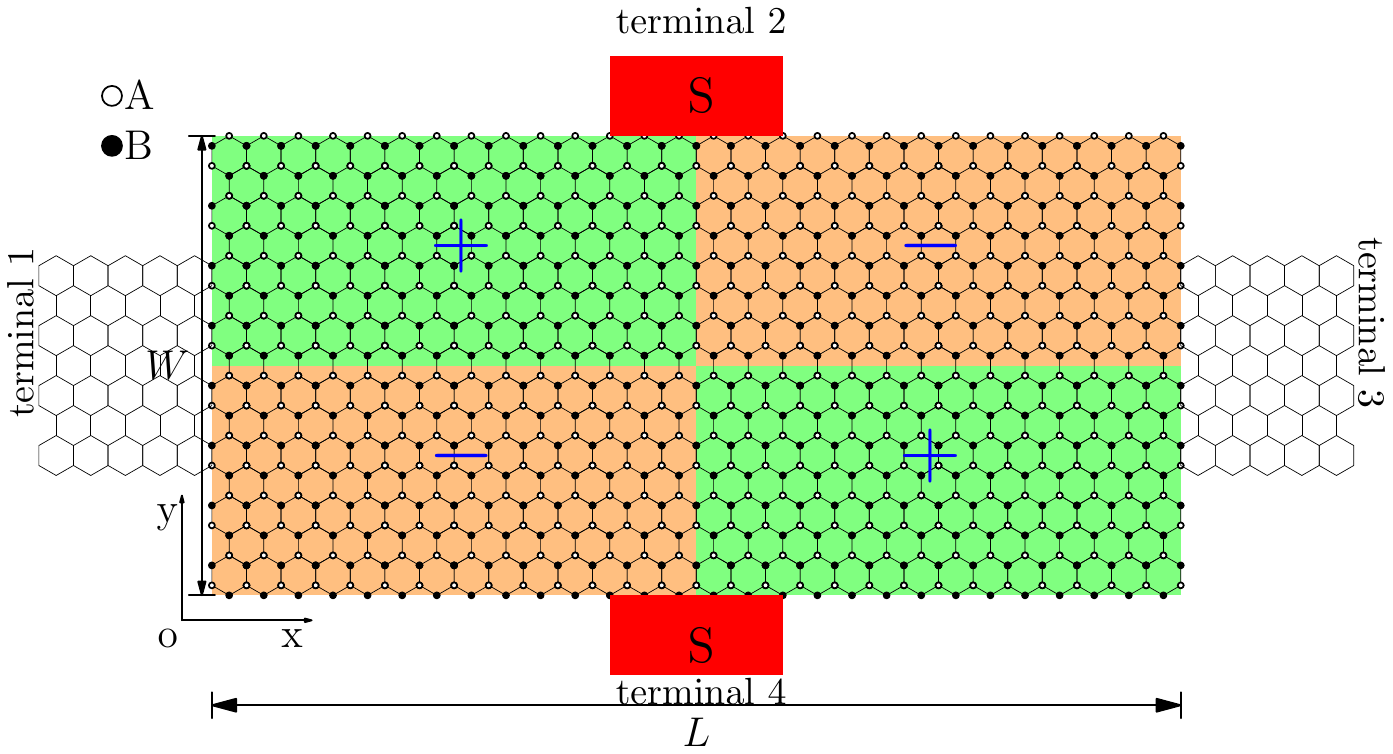}
  \caption{(Color online) The schematic diagram of the four-terminal device with the kink states. The terminal 1 and 3 are the graphene ribbons. The width and the length of the central region are $W=8$ and $L=28$, respectively. The terminal 2 and 4 are the superconductors. In the center region marked $'+'$(green), the on-site energy is $\varepsilon_{A}=V_{0}$ and $\varepsilon_{B}=-V_{0}$, where $V_{0}$ is called as the stagger energy. In the region marked $'-'$ (orange), $\varepsilon_{A}=-V_{0}$ and $\varepsilon_{B}=V_{0}$. The kink states are at the interface between two regions with the opposite stagger energy.}. \label{fig1}
\end{figure}

Topological insulator\cite{Moore2010} has been  one of the noticeable advanced materials due to the novel physical properties\cite{Hasan2010,Qi2010}, such as quantum Hall effect\cite{Laughlin1981,Thouless1982} and quantum spin Hall effect\cite{Kane2005a,Kane2005b}. It is of great significance to study the properties of topological insulator for developing a new generation of quantum components. Recently, the quantum anomalous Hall effect has been identified in the three-dimensional magnetic topological insulator\cite{Yu2010,Chang2013,Checkelsky2014,Kou2014}, which opens new possibilities for chiral-edge-state-based devices in zero external magnetic field\cite{Yasuda2017,Mogi2017}. The chiral states can also appear at the domain walls between two regions with the opposite Chern number\cite{Li2017}. For example, the graphene can be gapped by the sublattice symmetry breaking staggered on-site energy, which is showed in Fig. \ref{fig1}. There are one-dimensional states\cite{Qiao2011,June2011,Qiao2014,Cheng2018}, which are referred to as kink states below, presenting at the boundaries between regions with different quantized Hall conductances in the graphene nanoribbon.
A Majorana fermion, which is its own antiparticle, has the potential for quantum computing. Over the last two decades it has been realized that  Majorana fermions can emerge at zero-energy modes in topological superconductors\cite{Fu2009,Beenakker2015,Elliott2015}. After the chiral Majorana edge modes in the quantum anomalous Hall insulator-superconductor structure is realized in the experiment, there are more researches  focused on the topological superconductor device\cite{Xie2009,Hou2015,Song2017,Pacholski2018,wang2018,Wu2018,Wu2021}.

At the interface between a superconductor and a normal conductor, an incident electron from the normal conductor can be reflected as a hole, which is called Andreev reflection. When the bias voltage is lower than the superconductor gap, the conductance of the superconductor hybrid device is mainly determined by the progress of the Andreev reflection. Keeping to the time-reversal symmetry, the electron and the hole taking part in Andreev reflection come from the different valleys in the graphene-superconductor device\cite{Beenakker2006,Beenakker2008,wang2020}.

In this paper,  we calculate the transmission coefficients in a two-terminal device, where there are lattice deformations around the domain wall. The transmission coefficients are close to $1$ in the two-terminal device, which proves the robustness of the kink states\cite{Qiao2011,June2011,Qiao2014}. Then we study the Andreev reflection in the four-terminal device with the topological kink states showed in Fig. \ref{fig1}.  Through calculating the coefficients of the Andreev reflection in the four-terminal device, we find that the progress of the crossed Andreev reflection can be controlled by adjusting the on-site energy and the stagger energy.

The rest of this paper is arranged as follows. In Sec.\uppercase\expandafter {\romannumeral 2}, the model Hamiltonian for the system is presented and the formulas for calculating  the Andreev reflection coefficients are derived. Our main results are shown and discussed in Sec.\uppercase \expandafter {\romannumeral 3}. Finally, a brief conclusion is presented in Sec.\uppercase\expandafter {\romannumeral 4}.\\

\section{MODEL AND METHOD}
The four-terminal device with the kink states is showed in Fig. \ref{fig1}, where the terminal $1$ and $3$ are the normal graphene ribbons and the terminal $2$ and $4$ are superconductors. The center region is the colour region. The kink states are at the interface between the regions with different colour. Along the x direction and the y direction they are zigzag and armchair, respectively.  The total Hamiltonian of this junction can be represented as
\begin{equation}\
 H=H_{C}+H_{G}+H_{S}+H_{T},
\end{equation}
where$H_{C}$, $H_{G}$, $H_{S}$ and $H_{T}$ are the Hamiltonian of the center region, the graphene nanoribbons, the superconductor terminals and the coupling between the center region  and the superconductor terminals, respectively.

In the tight-binding representation,  $H_{C}$ and $H_{G}$ are given by
\begin{eqnarray}\label{HamiltonianTotal}
H_{C/G}=\sum_{n}\varepsilon_{0}a^{\dag}_{\emph{n}}a_{\emph{n}}-\sum_{<m,n>}[ta_{\emph{n}}^{\dagger}a_{\emph{m}}+H.c.],\qquad    
\end{eqnarray}
where $a_{\emph{n}}^{\dagger}$ and $a_{\emph{n}}$ are the creation operator and annihilation operators of the $n$th point. $\varepsilon_{0}$ is the on-site energy in the graphene terminals and the center region. In the center region, $\varepsilon_{0}=\pm V_{0}$ for the different partition, where $V_{0}$ is the stagger energy of A and B sublattice. The second term in Eq. (\ref{HamiltonianTotal}) stands for the nearest-neighbor hopping Hamiltonian. Considering that the center region is directly coupled to the superconductor terminals, we use the BCS Hamiltonian for the superconductor terminals described by a continuum model,
\begin{eqnarray}
H_{S}=\sum_{\textbf{\emph{k}},\sigma}\varepsilon_{\textbf{\emph{k}}}C^{\dagger}_{\textbf{\emph{k}}\sigma}C_{\textbf{\emph{k}}\sigma}
+\sum_{\textbf{\emph{k}}}(\Delta C_{\textbf{\emph{k}}\downarrow}C_{-\textbf{\emph{k}}\uparrow}
+\Delta^{*}C^{\dagger}_{-\textbf{\emph{k}}\uparrow}C^{\dagger}_{\textbf{\emph{k}}\downarrow}),\qquad
\end{eqnarray}
where $\Delta=\Delta_{s}e^{\mathrm{i}\theta}$. Here $\Delta_{s}$ is the superconductor gap and $\theta$ is the
superconductor phase. The coupling between superconductor terminals and graphene is described by
\begin{eqnarray}
H_{T}=-\sum_{\emph{n},\sigma}\emph{t}_ca^{\dagger}_{\emph{n},\sigma}C_{\sigma}(x_{\emph{i}})+H.c.
\end{eqnarray}
Here $x_{\emph{i}}$ and $n$ represent the positions of the coupling atoms on the interface of superconductor and the center region, and
$C_{\sigma}(x)=\sum_{\emph{k}_{x}}e^{\mathrm{i}k_{x}x}C_{\textbf{\emph{k}},\sigma}$ is the annihilation operator in real space. Note that $\sigma$ represents the spin index and $\emph{t}_c$ is the coupling strength between graphene and superconductor terminals.

We now turn to analyze the process that an incident electron from the graphene terminal is reflected into a hole with a Cooper pair emerging in the superconductor terminal. Using nonequilibrium Green's function method, we can calculate the retarded and advanced Green's function $G^{r}(E)=[G^{a}]^{\dagger}=1/(EI-H_{C}-\sum_{\alpha}\mathbf{\Sigma}^{r}_{\alpha})$, where $H_{C}$ is the Hamiltonian of the center region in the Nambu representation and $I$ is the unit matrix with the same dimension as $H_{C}$. $\mathbf{\Sigma}_{\alpha}^{r}=t_c g_{\alpha}^{r}(E)t_c$ is the retarded self-energy due to the coupling to the terminal $\alpha$,  where $g_{\alpha}^{r}(E)$  is the surface Green's function of the terminal $\alpha$. We can numerically calculate the surface Green's function of the graphene terminals. For superconductor terminals, the surface Green's function\cite{Xie2009,wang2018,Song2017} in real space is
\begin{eqnarray}\label{1}
 g_{\alpha,ij}^{r}(E)=i\pi\rho\beta(E)J_{0}[k_{f}(x_{i}-x_{j})]\left(
\begin{array}{cc}
  1 & \Delta/E \\
  \Delta^{*}/E & 1 \\
\end{array}
\right),\qquad
\end{eqnarray}
where $\alpha=2,4$ and $\rho$ is the density of normal electron states. $J_{0}[k_{f}(x_{i}-x_{j})]$ is the Bessel function of the first kind with the Fermi wavevector $k_{f}$. $\beta(E)=-iE/\sqrt{\Delta_{s}^{2}-E^2}$ for $|E|<\Delta_{s}$ and $\beta(E)=|E|/\sqrt{E^2-\Delta_{s}^{2}}$ for $|E|>\Delta_{s}$.

The Andreev reflection  coefficients for the incident electron coming from the graphene terminal $1$ can  be obtained\cite{Xie2009} by
\begin{eqnarray}\label{AndRfC}
&T_{A,11}(E)=\mathrm{Tr}\{\Gamma_{1,\uparrow\uparrow}G^{r}_{\uparrow\downarrow}\Gamma_{1,\downarrow\downarrow}G^{a}_{\downarrow\uparrow}\},\nonumber\\
&T_{A,13}(E)=\mathrm{Tr}\{\Gamma_{1,\uparrow\uparrow}G^{r}_{\uparrow\downarrow}\Gamma_{3,\downarrow\downarrow}G^{a}_{\downarrow\uparrow}\},
\end{eqnarray}
where the subscripts $\uparrow\uparrow$, $\downarrow\downarrow$, $\uparrow\downarrow$ and $\downarrow\uparrow$ represent the 11, 22, 12 and 21 matrix elements, respectively, in the Nambu representation. The linewidth function $\Gamma_{\alpha}$ is defined with the aid of self-energy as $\Gamma_{\alpha}=i[\mathbf{\Sigma}_{\alpha}^{r}-(\mathbf{\Sigma}_{\alpha}^{r})^{\dag}]$. $T_{A,11}$ and $T_{A,13}$ represent the coefficients of the local Andreev  reflection and the  crossed Andreev reflection, respectively.  Because the Andreev reflection from an electron to a hole is equivalent to that from a hole to an electron under particle-hole symmetry, in this work we only consider the Andreev reflection, where an incident electron is reflected as a hole. \\

\section{NUMERICAL RESULTS AND ANALYSIS}
In numerical calculations, we set the nearest-neighbour hopping energy $t=2.75 eV$.  The length of the nearest-neighbor C-C bond is set to be $a_{0}=0.142$ $\mathrm{nm}$ as in a real graphene sample. The superconductor gap is set to be $\Delta_{s}=0.02t$ and the Fermi wavevector $k_{f}=10$ $\mathrm{nm^{-1}}$. For the convenience of discussing the influence of the kink states on Andreev reflection,  The Fermi energy $E_{\textit{f}}$  is set to be zeros in our calculations for the convenient.

\begin{figure}[b]
\centering
\includegraphics[width=1.0\columnwidth,angle=0]{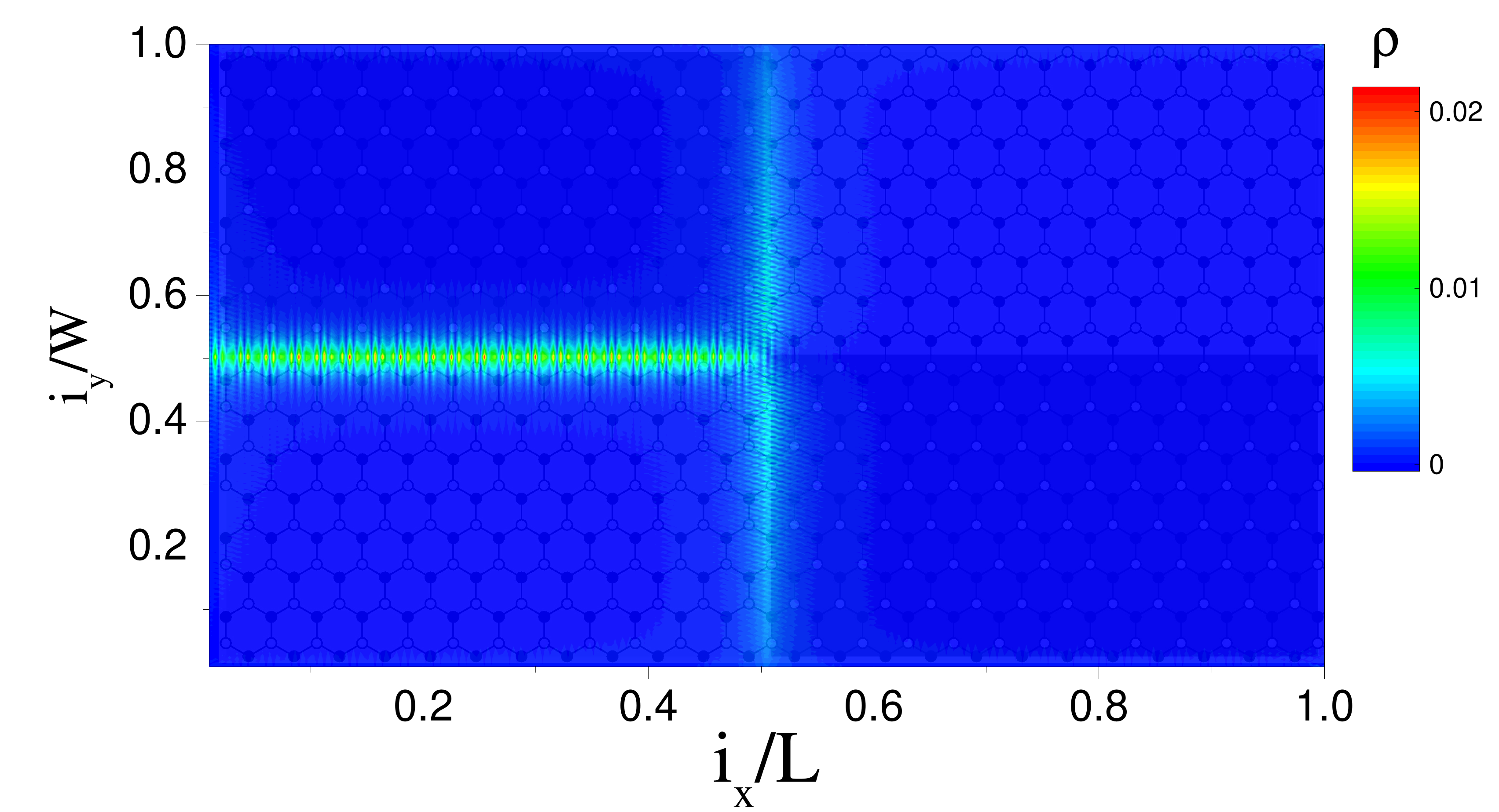}
\caption{(Color online) The local density of states distribution  for modes incident from the terminal $1$. All of the terminals are the graphene nanoribbons. The width and the length of the central region are $W=80$ and $L=120$, respectively.} \label{fig2}
\end{figure}

Fig. \ref{fig2} shows the local density of states contribution of scattering states injected from the terminal $1$ when the four terminals connected with the center region are the graphene nanoribbon. From Fig. \ref{fig2}, we can see that in the center region the electrons can only travel along the interface between two regions with the opposite value of stagger energy.It is clear that when the electrons injected from terminal $1$,  they can only travel along the kink states  into the terminal $2$ and the terminal $4$. The propagation to terminal $3$ is forbidden.  It shows that an electronic beam splitter is created by this design.  At the intersection point propagation in the forward direction from the terminal $1$ to the terminal $3$ is forbidden because of  the reversed pseudospin in the direction of propagation along the $x$ direction\cite{Qiao2011}, which is explained in the energy band below.

\begin{figure}[tbp]
\centering
\includegraphics[width=1.0\columnwidth,angle=0]{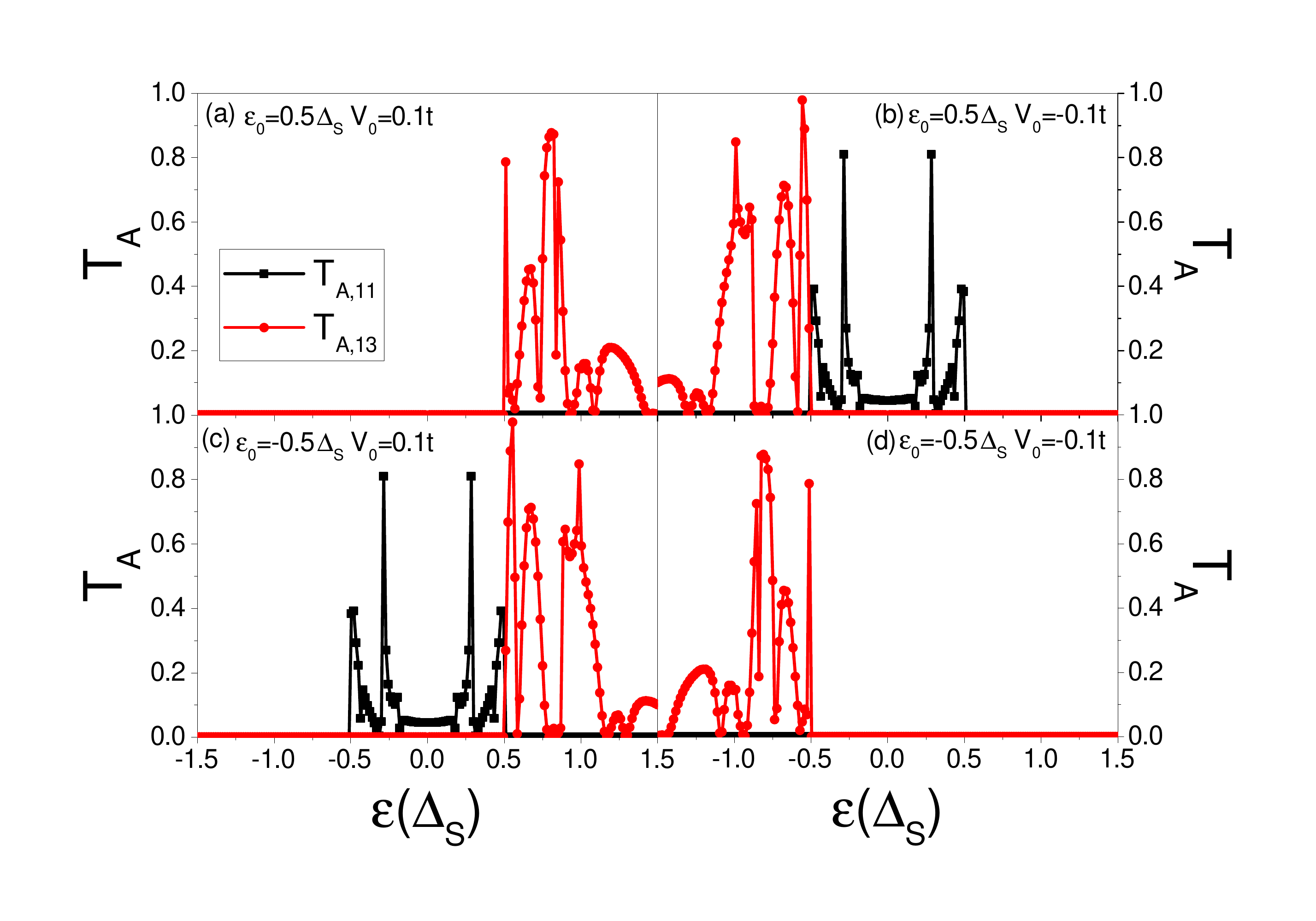}
\caption{(Color online)  $T_A$ vs the incident energy $\varepsilon$ in the three-terminal device.  The on-site energy of the garphene terminal is $\varepsilon_{0}=0.5\Delta_{s}$ in (a) and (b) and $\varepsilon_{0}=-0.5\Delta_{s}$ in (c) and (d). The stagger energy is $V_{0}=0.1t=50\Delta_{s}$ in (a) and (c) and $V_{0}=-0.1t=-50\Delta_{s}$  in (b) and (d). The width and the length of the central region are $W=80$ and $L=120$, respectively.} \label{fig3}
\end{figure}

It is certified that how the electrons travel along the kink states in the four-terminal device in Fig. \ref{fig2}. Next we want to know the influences of the kink states on the Andreev reflection. The superconductor gap is set to be $\Delta_{s}=0.02t$, which is more less than the stagger energy $|V_{0}|=0.1t$, so that only kink states take part in the progress of Andreev reflection.  Firstly, we study a three-terminal superconductor device, where only one superconductor terminal $2$ or $4$ is  retained in the four-terminal device, which is showed in Fig. \ref{fig1}.  The coefficients of the local Andreev reflection $T_{A,11}$ and the crossed Andreev reflection $T_{A,13}$ varying with the incident energy $\varepsilon$ in the three-terminal device are presented in Fig. \ref{fig3}.

In Fig. \ref{fig3} (a), when $\varepsilon_{0}=0.5\Delta_{s}$ and $V_{0}=0.1t$, the coefficient of local Andreev reflection is $T_{A,11}=0$, so the local Andreev reflection is suppressed. The coefficient of the crossed Andreev reflection is zero for $-1.5\Delta_{s}<\varepsilon<0.5\Delta_{s}$. When the incident energy  increases to  $\varepsilon=0.5\Delta_{s}$, $T_{A,13}$ changes from $0$ to $0.8$ quickly, then oscillates with $\varepsilon$ increasing. $T_{A,13}$ reaches the peak value about $0.9$ when $\varepsilon$ increases to $0.85\Delta_{s}$. With $\varepsilon$ increasing more than $0.9\Delta_{s}$, $T_{A,13}$ decreases quickly. Fixing the on-site energy $\varepsilon_{0}=0.5\Delta_{s}$ and adjusting the stagger energy to $V_{0}=-0.1t$ in Fig. \ref{fig3} (b),  we can see that there is the local Andreev reflection arising for $-0.5\Delta_{s}<\varepsilon<0.5\Delta_{s}$. $T_{A,11}$ is symmetrical about the point $\varepsilon=0$ and reaches the peak value $0.8$ at the points of $|\varepsilon|=0.25\Delta_{s}$. $T_{A,13}$ is zero for $-0.5\Delta_{s}<\varepsilon<1.5\Delta_{s}$. When the incident energy decreases from $-0.5\Delta_{s}$, $T_{A,13}$ increases to the peak value $1.0$ quickly and oscillates with $\varepsilon$ decreasing. $T_{A,13}$ reaches the second highest peak value about $0.85$ at $\varepsilon=-1.0\Delta_{s}$, then decreases quickly with $\varepsilon$ decreasing less than  $-1.0\Delta_{s}$. When the stagger energy is set to be $V_{0}=1.0t$ and the on-site energy is $-0.5\Delta_{s}$,  the curves in Fig. \ref{fig3} (c) are the same as the  mirror image of that in Fig. \ref{fig3} (b). Setting  $V_{0}=-1.0t$ and  $\varepsilon_{0}=-0.5\Delta_{s}$, we find that  the local Andreev reflection is suppressed again and the curve of $T_{A,13}$ in Fig. \ref{fig3} (d) is the same as  the mirror image of that in Fig. \ref{fig3} (a).

In Fig. \ref{fig3}, $\varepsilon=|\varepsilon_{0}|$ are a critical points between $T_{A,11}$ and $T_{A,13}$. In Fig. \ref{fig3} (a) and (d) the local Andreev reflection is suppressed and the crossed Andreev reflection takes place for $\varepsilon>|\varepsilon_{0}|$ in Fig. \ref{fig3} (a) and $\varepsilon<-|\varepsilon_{0}|$ in Fig. \ref{fig3} (d). In Fig. \ref{fig3} (b) and (c),  $T_{A,11}$ is nonzero for $\varepsilon<|\varepsilon_{0}|$. $T_{A,13}$ is zeros for $\varepsilon>-|\varepsilon_{0}|$  and $\varepsilon<|\varepsilon_{0}|$ in Fig. \ref{fig3} (b) and (c), respectively.  When the sign of the on-site energy $\varepsilon_{0}$ and the stagger energy $V_{0}$ are set to be same, the local Andreev reflection is suppressed. To sum up,  the progress of Andreev reflection can be controlled by adjusting the on-site energy and the stagger energy in the device with kink states.

\begin{figure}[b]
\includegraphics[width=1.0\columnwidth,angle=0]{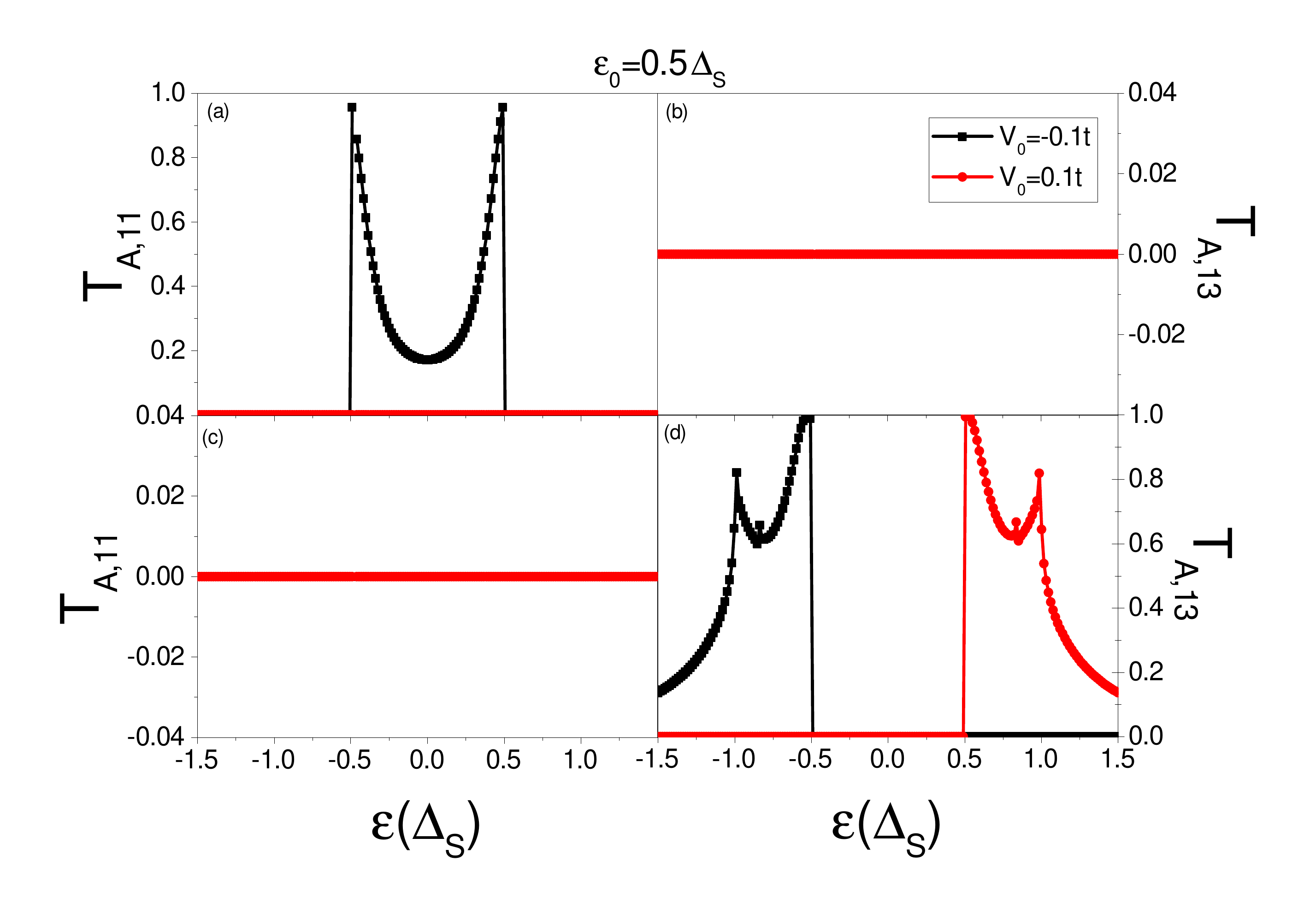}\\
\caption{(Color online) The coefficients of the local Andreev reflection $T_{A,11}$ and the crossed Andreev reflection $T_{A,13}$ vs the incident energy $E$ in the four-terminal graphene-superconductor junction.  The phase difference between two superconductor terminals is $\triangle\phi=0$ in (a) and (b) and $\triangle\phi=\pi$ in (c) and (d). } \label{fig4}
\end{figure}

Then when we set the on-site energy $\varepsilon_{0}=0.5\Delta_{s}$, the coefficients of Andreev reflection in the four-terminal device are showed in Fig. \ref{fig4} for $V_{0}=\pm 0.1t$. In the previous works, Andreev reflection can be adjusted by the phase difference of the superconductor terminals $\triangle\phi$ in the four-terminal device\cite{cheng2009,xing2011}. The local Andreev reflection is suppressed when $\triangle\phi=\pi$, conversely, the crossed Andreev reflection is suppressed when  $\triangle\phi=0$.  In Fig. \ref{fig4} (b) and (c), it is clear that the crossed Andreev reflection $T_{A,13}$ is suppressed for $\triangle\phi=0$ and the local Andreev reflection $T_{A,11}$ is suppressed for $\triangle\phi=\pi$. These results are in accord with the previous work\cite{cheng2009,cheng2015}. It is important to note that in Fig. \ref{fig4} (a) for $V_{0}=0.1t$ $T_{A,11} $ is always zero when $\triangle\phi=0$,  which is different from the results of the previous works. That is to say when $\triangle\phi=0$ the local Andreev reflection can be suppressed by adjusting the stagger energy. In Fig. \ref{fig4} (d), for $\varepsilon_{0}=\pm 0.5\Delta_{s}$, when $V_{0}=0.1t$, $T_{A,13}$ keeps zero for $\varepsilon<|\varepsilon_{0}|$;  on the contrary, when $V_{0}=-0.1t$,  $T_{A,13}$ keeps zero for $\varepsilon>-|\varepsilon_{0}|$.
\begin{figure}[tbp]
\centering
\includegraphics[width=1.0\columnwidth,angle=0]{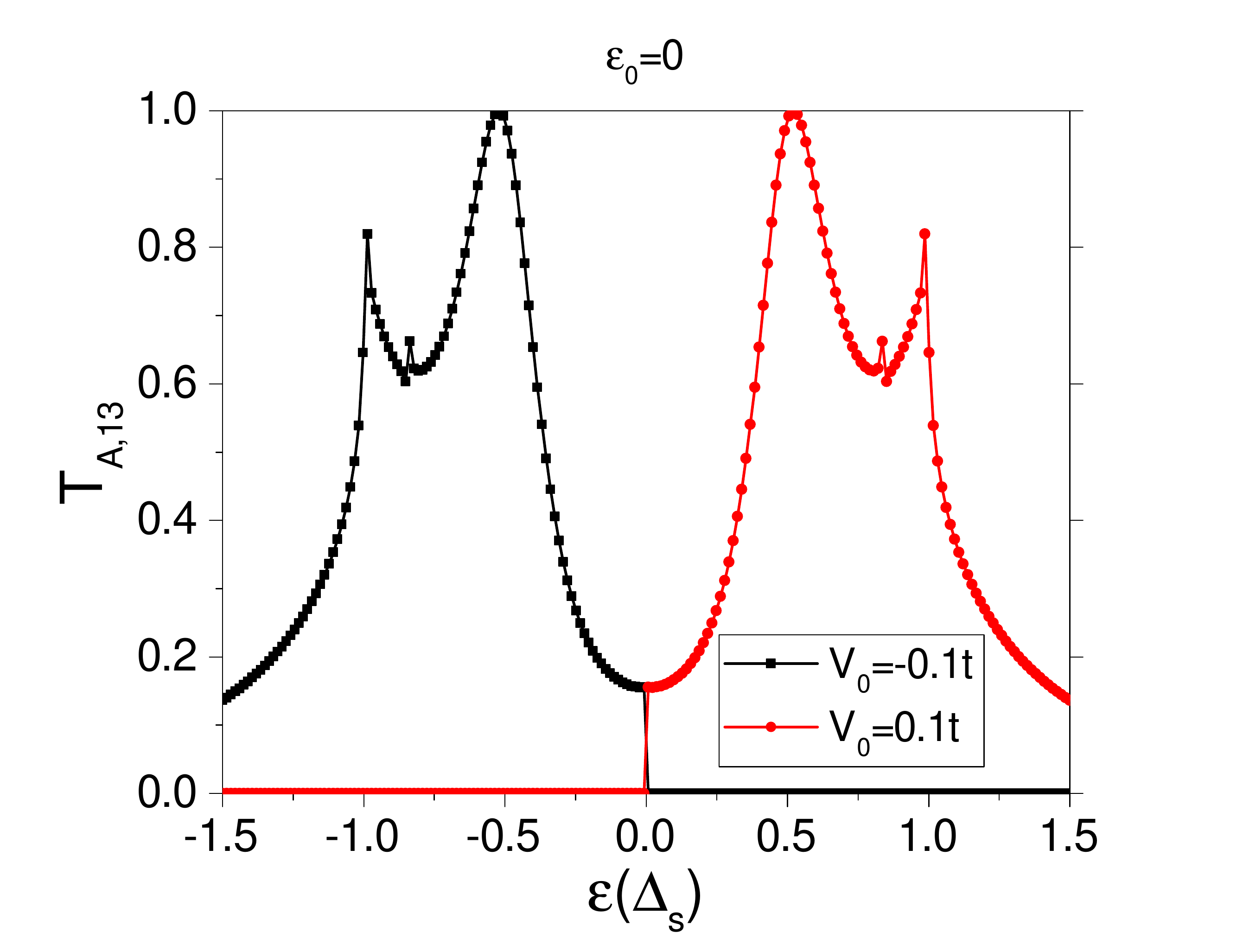}
\caption{(Color online)  $T_A$ vs the incident energy $\varepsilon$ in the four-terminal device. The phase difference between  superconductors  is $\triangle\phi=\pi$. } \label{fig5}
\end{figure}

\begin{figure}[t]
\centering
\includegraphics[width=1.0\columnwidth,angle=0]{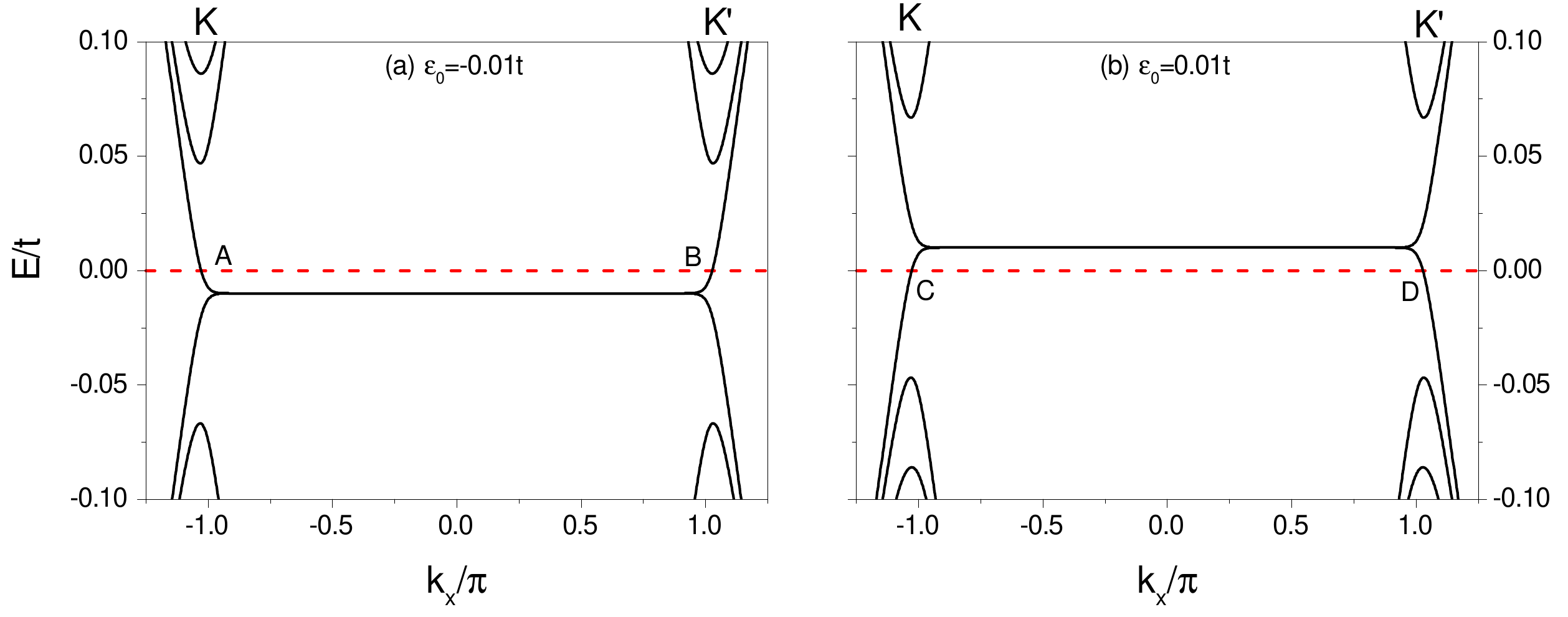}
\includegraphics[width=1.0\columnwidth,angle=0]{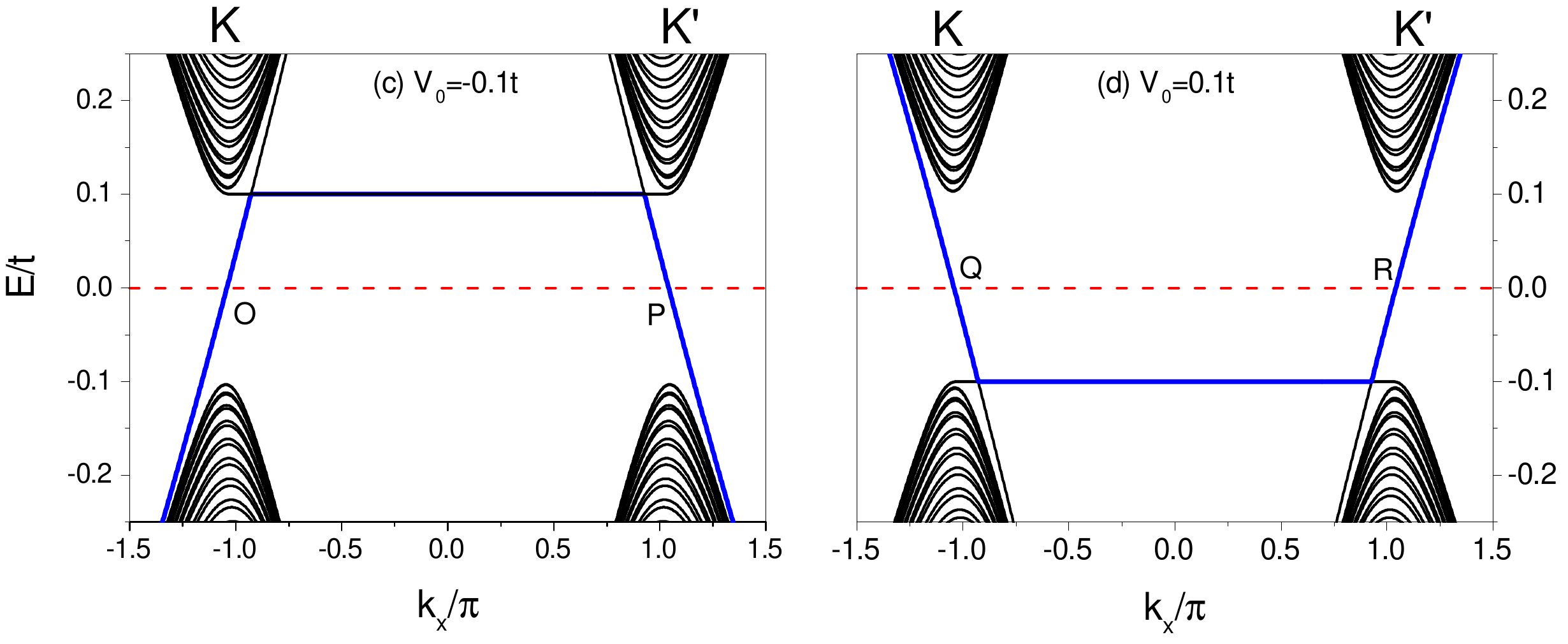}
\caption{(Color online) (a) and (b) are the band structures of the graphene terminals with  different $E_{\textit{f}}$. (c) and (d) are the band structures of center region with different stagger energy $V_{0}$ in the x direction. The lines in blue represent the kink states.} \label{fig6}
\end{figure}

After that, the on-site energy is set to be zero.  In Fig. \ref{fig5}, as the incident energy changing the coefficients of the crossed Andreev reflection $T_{A,13}$ in the four-terminal device are showed for the on-site energy $\varepsilon_{0}=0$ with $\triangle\phi=\pi$. When $V_{0}=0.1t$, the coefficient is zero for the incident energy $\varepsilon<0$. Increasing the incident energy, the coefficient changes from $0$ to $0.2$ rapidly. When $\varepsilon$ comes  up to $0.5\Delta_{s}$, the coefficient reaches the peak. Continuing increasing $\varepsilon$, the coefficient decreases gradually. When $V_{0}=-0.1t$, $T_{A,13}$ is zero for $\varepsilon>0$ and reaches the peak at $\varepsilon=-0.5\Delta_{s}$.  The peak value of the coefficients can reach  $1$ at $\varepsilon=|0.5\Delta_{s}|$, namely all of the reflected hole can transport into terminal $3$ in this device in some conditions. To explain this case we plot the  energy band structures of the terminal and the center region in Fig. \ref{fig6}.

\begin{figure}[t]
\centering
\includegraphics[width=1.0\columnwidth,angle=0]{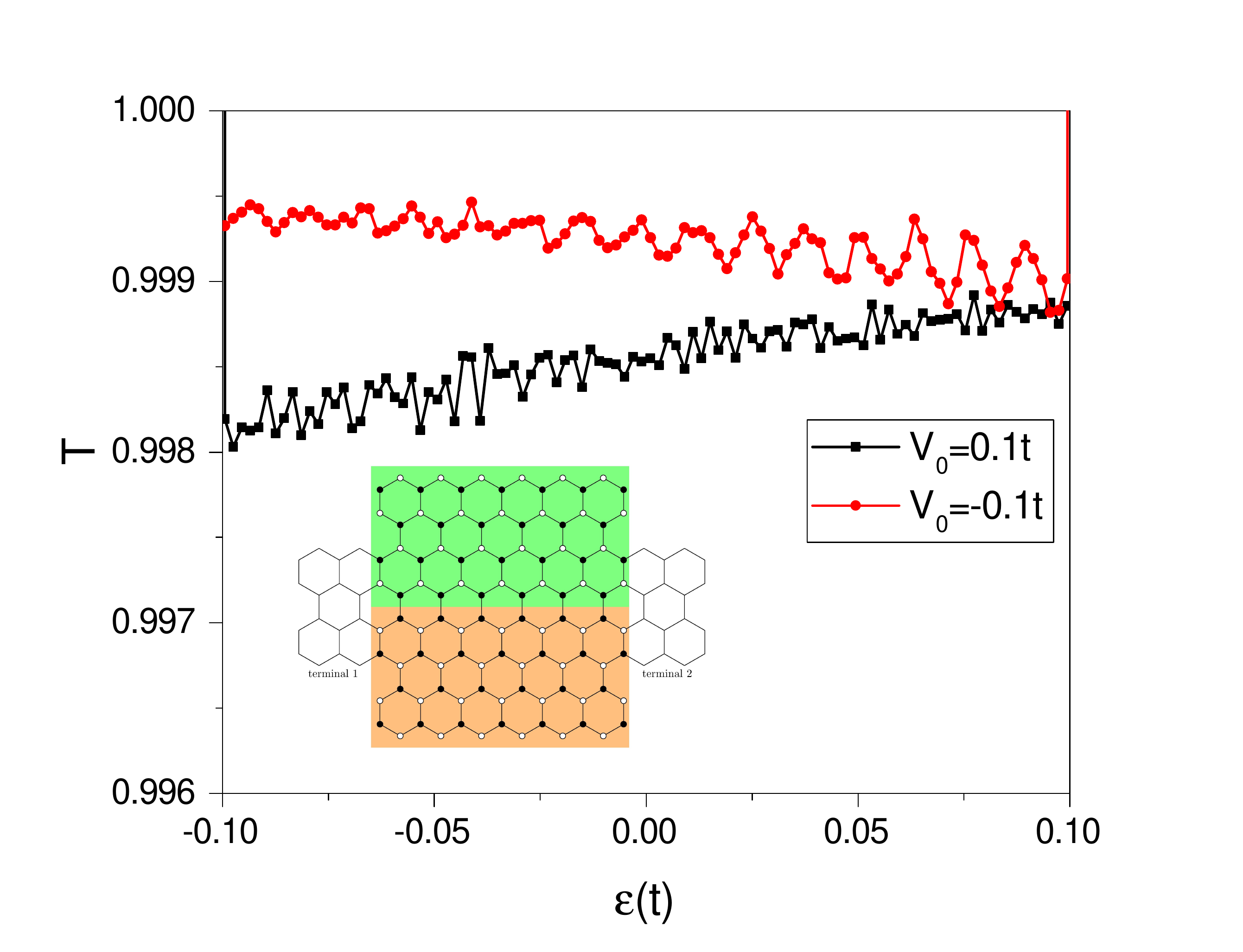}\\
\caption{(Color online)  The transmission coefficients vs the incident energy $E$ in the two-terminal junction with zigzag ribbon under the lattice deformation. The width and the length of the central region are $W=80$ and $L=60$, respectively. } \label{fig7}
\end{figure}

Fig. \ref{fig6} (a) and (b) are the energy band structures of the terminal $1$ and $3$ with $\varepsilon_{0}=\pm 0.5\Delta_{s}=\pm 0.01t$. Fig. \ref{fig6} (c) and (d) are the energy bands of the center region with opposite stagger energy, respectively. In Fig. \ref{fig7} (a), the point A and B are in the conduction band. The point A is in the valley $K$ and  represents the state of outgoing. The point B is in the valley $K'$  and represents the state of incoming. In Fig. \ref{fig6} (b), the point C and D are in valence band. The point C in the valley K represents the state of incoming and the point D in valley $K'$ represents the state of outgoing. Similarly, in Fig. \ref{fig6} (c) and (d) the point O and R (P and Q) represent the state of incoming (outgoing), which belong to the valley $K$ and $K'$, respectively.

When the incident electrons from terminal $1$ travel into the center region,  which come from the point $B$ in the conduction band,  they can  transport into the point  $O$ or the  point $R$ due to the conservation of the momentum.  Beside the conservation of the momentum, the travel of the electron must make the pseudospin invariant. So the transport from the point $B$ to the point $O$ is forbidden and the incident electrons from the point $B$ can only travel into the point $R$.  In other words, for the low-potential regions, when the incident electrons coming from the valley $K$ (the conduction band) or $K'$ (the valence band) travel into the center region, they should travel into the same valley in the band of center region with the kink states, otherwise, propagation is forbidden. It can explain why the propagation is forbidden in Fig. \ref{fig2}. In Fig. \ref{fig5}, when the incident energy $\varepsilon>\varepsilon_{0}$, the electrons come from the point $B$ in the conduction band.  So the incident electrons can only travel into point R for $V_{0}=0.1t$ and the propagation of the electrons is forbidden for $\varepsilon<\varepsilon_{0}$ in this case.

There is another question to answer.  When $\varepsilon_{0}=0$ the local Andreev reflection is suppressed and $T_{A,11}$ is zeros for $\Delta\phi=\pi$ which is beneficial for the crossed Andreev reflection in the four-terminal device. However, we find that for $\Delta\phi=0$, which is beneficial for the local Andreev reflection, the coefficient $T_{A,11}$ is always zeros when $\varepsilon_{0}=0$.  The figure is not showed in the paper.

What is the reason of this case?   From Fig. \ref{fig2}, we know that the incident electrons with $\varepsilon$ travel along the kink states to the interfaces between the center region and the superconductors. The reflected holes with $-\varepsilon$  must travel along the kink states, too. From our discussions, we know that due to the conservation of the momentum and the pseudospin, propagation of the incident electrons can be  forbidden in some conditions.  The same reason applies to  propagation of the reflected holes. For example, when $\varepsilon_{0}=0$,  the incident electrons from the valley $K$ can travel into the center region along the zigzag kink states for $\varepsilon>\varepsilon_{0}$ and $V_{0}=0.1t$. As the incident electrons flow into the superconductor along the armchair kink states, the reflected holes with $-\varepsilon$ travel back to the center region along the armchair kink states. At the intersection, along the zigzag kink states the reflected holes flowing in the  $-x$ direction  belongs to the  valley $K'$ and can not flow into the terminal $1$. So the reflected holes flowing in the $x$ direction, which belong to the valley $K$, can only travel into the terminal $3$.

The lattice deformations around the interface are common in the real device. There is a question how the lattice deformations influence the kink states. The two-terminal device with the kink states is showed in the inset of Fig. \ref{fig7}. Considering the deformations of the lattices near the interface, we calculate the  transmission coefficients for $V_{0}=\pm 0.1t$ to answer the question. The transmission coefficient is $T=\mathrm{Tr}\{\Gamma_{L}G^{r}\Gamma_{R}G^{a}\}$, where $L$ and $R$ represents the terminal $1$ and $2$ in the two-terminal device, respectively. We set the deformation of the lattice to be a random value in the range of $[0,0.2a]$. The processes of the calculation are cycled for $500$ times and the average values are showed in Fig. \ref{fig7}. Due to the lattice deformation, there should be scattering effect at the interface, the transport coefficients are a little less than $1$ for  the stagger energy $V_{0}=0.1t$ or $V_{0}=-0.1t$. From the Fig. \ref{fig7}, we can see that the influence of the lattice deformation on transport is very slight, the minimum value of the transport coefficients is about $0.998$, which is very close to $1$. The kink states are robust, so it is a good choice for designing quantum devices.

\section{CONCLUSIONS}
We study the influence of the lattice deformations on the kink states in a two-terminal device. Through our research, we certify that the lattice deformations could weaken the  electron  transport along the kink states, but the influence is very small. So the kink states are robustness under the lattice deformations.

In the three-terminal device, we find that through adjusting  the on-site energy and the stagger energy, for example  $\varepsilon_{0}=0.5\Delta_{s}$ and $V_{0}=1.0t$, the local Andreev reflection  can be completely suppressed. In the four-terminal device, due to the conservation of the momentum and the pseudospin,  both the local Andreev reflection and the crossed Andreev reflection can be suppressed when $\triangle\phi=0$ and $V_{0}=0.1t$. This results show us that  the progress of Andreev reflection can be controlled by adjusting the on-site energy $\varepsilon_{0}$ and the stagger energy $V_{0}$ in the multi-terminal device with the kink states. It should give us some new ideas on the research and development of the quantum device.

\begin{acknowledgments}
This work was supported by the National Natural Science Foundation of China (Grant Nos. 11874139, 11474085), the Natural Science Foundation of Hebei (Grant Nos. A2017205108, A2019205190), the youth talent support program of Hebei education department (Grant No. BJ2014038), the Outstanding Youth Foundation of HBTU (Grant No. L2016J01), the science program of Shijiazhuang (Grant No. 201790741), the youth talent support program of Hebei Province, and Doctoral Research Foundation of Shijiazhuang College (Grant No. 20BS017).\\
\end{acknowledgments}



\begin{thebibliography}{99}
\bibitem[*]{} Corresponding authors: \\1102029@sjzc.edu.cn; yxli@hebtu.edu.cn

\bibitem[1]{Moore2010}J. E. Moore, Nature. {\bf 464}, 194 (2010).

\bibitem[2]{Hasan2010}M. Z. Hasan and C. L. Kane, Rev. Mod. Phys. {\bf 82}, 3045 (2010).

\bibitem[3]{Qi2010}X. L. Qi and S. C. Zhang, Rev. Mod. Phys. {\bf 83}, 1057 (2010).

\bibitem[4]{Laughlin1981}R. B. Laughlin,, Phys. Rev. B. {\bf 23}, 5632 (1981).

\bibitem[5]{Thouless1982} D. J. Thouless, M. Kohmoto, M. P. Nightingale, and M. den Nijs, Phys. Rev. Lett. {\bf 49}, 405 (1982).

\bibitem[6]{Kane2005a}C. L. Kane, and E. J. Mele, Phys. Rev. Lett. {\bf 95}, 226801 (2005).

\bibitem[7]{Kane2005b}C. L. Kane, and E. J. Mele, $Z_{2}$ , Rev. Lett. {\bf 95}, 146802 (2005).

\bibitem[8]{Yu2010} R. Yu, W. Zhang, H. J. Zhang, S. C. Zhang, X. Dai, and Z. Fang, Science {\bf 329}, 61 (2010).

\bibitem[9]{Chang2013} C. Z. Chang, J. S. Zhang, X. Feng, J. Shen, Z. C. Zhang, M. H. Guo, K. Li, Y. B.o Ou, P. Wei, L. L.  Wang, Z. Q. Ji, Y. Feng, S. H. Ji, X. Chen, J. F. Jia, X. Dai, Z. Fang, S. C. Zhang, K. He, Y. Y. Wang, L. Lu, X. C. Ma, and Q. K. Xue, Science {\bf 340}, 167 (2013).

\bibitem[10]{Checkelsky2014} J. G. Checkelsky, R. Yoshimi, A. Tsukazaki, K. S. Takahashi, Y. Kozuka, J. Falson, M. Kawasaki, and Y. Tokura, Nature Physics {\bf 10}, 731 (2014).

\bibitem[11]{Kou2014} X. F. Kou, S. T. Guo, Y. B. Fan, L. Pan, M. R. Lang, Y. Jiang, Q.M. Shao, T. X. Nie, K. Murata, J. S. Tang, Y. Wang, L. He, T. K. Lee, W. L. Lee, and K. L. Wang, Phys. Rev. Lett. {\bf 113}, 137201 (2014).

\bibitem[12]{Yasuda2017} K. Yasuda, M. Mogi, R. Yoshimi, A. Tsukazaki, K. S. Takahashi, M. Kawasaki, F. Kagawa, and Y. Tokura, Science {\bf 358}, 1311 (2017).

\bibitem[13]{Mogi2017} M. Mogi, M. Kawamura, R. Yoshimi, A. Tsukazaki, Y. Kozuka, N. Shirakawa, K. S. Takahashi, M. Kawasaki,and  Y. Tokura, Nat. Mater. {\bf 16}, 516 (2017).

\bibitem[14]{Li2017} J. Li, R. X. Zhang, Z. X. Yin, J. X. Zhang, K. Watanabe, T. Taniguchi, C. X. Liu, and J. Zhu,  Science {\bf 362}, 1149 (2018).

\bibitem[15]{Qiao2011} Z. H. Qiao, J. Jung, Q. Niu, and A. H. MacDonald, Nano lett. {\bf 11}, 3453 (2011).

\bibitem[16]{June2011} J. Jung, F. Zhang, Z. H. Qiao, and A. H. MacDonald, Phys. Rev. B {\bf 84}, 075418 (2011).

\bibitem[17]{Qiao2014} Z. H. Qiao, J. Jung, C. W. Lin, Y. F.  Ren, A. H. MacDonald and Q. Niu, Phys. Rev. B {\bf 112}, 206601 (2014).

\bibitem[18]{Cheng2018} S. G. Cheng, H. W. Liu, H. Jiang, Q. F. Sun, and X. C. Xie,  Phys. Rev. Lett. {\bf 121}, 156801 (2018).

\bibitem[19]{Fu2009}L. Fu and C. L. Kane,  Phys. Rev. Lett.  {\bf 102}, 216403 (2009).

\bibitem[20]{Beenakker2015}C. W. J. Beenakker, Rev. Mod. Phys. {\bf 87}, 1037 (2015).

\bibitem[21]{Elliott2015}S.  R. Elliott, and M. Franz, Rev. Mod. Phys. {\bf 87}, 137 (2015).

\bibitem[22]{He2017}Q. L. He, L. Pan, A. L. Stern, E. C. Burks, X. Y. Che,G. Yin, J. Wang, B. Lian, Q. Zhou, E. S. Choi, K. Murata, X. F. Kou, Z. J. Chen, T. X. Nie,Q. M. Shao, Y. B. Fan, S. C. Zhang, K.  Liu, J. Xia,and  K. L. Wang, Science. {\bf 357}, 294 (2017).

\bibitem[23]{Xie2009} Q. F. Sun and X. C. Xie, J. Phys.: Condens. Matter {\bf 21}, 344204 (2009).

\bibitem[24]{Hou2015} Z. Hou, and Q. F. Sun, Phys. Rev. B {\bf 96}, 155305 (2017).

\bibitem[25]{Song2017} J. Liu, H. W. Liu, J. T. Song, Q. F. Sun and X. C. Xie, Phys. Rev. B {\bf 96}, 045401 (2017).

\bibitem[26]{Pacholski2018} M. J. Pacholski, C. W. J. Beenakker, and I. Adagideli, Phys. Rev. Lett {\bf 121}, 037701 (2018).

\bibitem[27]{wang2018} C. Wang, Y. L. Zou,  J. T. Song and Y. X. Li, Phys. Rev. B {\bf 98}, 035403 (2018).

\bibitem[28]{Wu2018} Hai-Bin Wu, Ying-Tao Zhang and Jian-Jun Liu, Journal of Applied Physics {\bf 124}, 084301 (2018).

\bibitem[29]{Wu2021} Hai-Bin Wu and Jian-Jun Liu, Phys. Rev. B {\bf 103}, 115430 (2021).

\bibitem[30]{Beenakker2006} C. W. J. Beenakker, Phys. Rev. Lett. {\bf 97}, 067007 (2006).

\bibitem[31]{Beenakker2008} C. W. J. Beenakker, Rev. Mod. Phys. {\bf 80}, 1337 (2006).

\bibitem[32]{wang2020} C. Wang, L. Zhang, P. P. Zhang, J. T. Song and Y. X. Li, Phys. Rev. B {\bf 101}, 045407 (2020).

\bibitem[33]{cheng2009} S. G. Cheng, Y. Xing, J. Wang, and Q. F. Sun, Phys. Rev. B {\bf 103}, 167003 (2009)

\bibitem[34]{xing2011}  Y. X. Xing, J. Wang, and Q. F. Sun, Phys. Rev. B {\bf 83}, 205418 (2011)

\bibitem[35]{cheng2015} F. Li and S. G. Cheng, J. Phys.: Condens. Matter {\bf 27}, 125002 (2015).

\end{thebibliography}
\end{document}